\documentclass[sigconf, screen,nonacm]{acmart}
\AtBeginDocument{%
  }

\setcopyright{none}

\usepackage[capitalise]{cleveref}
\usepackage{acronym}
\usepackage{placeins}
\usepackage{subcaption}
\usepackage{multirow}
\usepackage[inline]{enumitem}
\usepackage{fontawesome5}

\acrodef{rs}[RS]{recommender systems}
\acrodef{ra}[RA]{recommen\-dation algorithm}
\acrodef{ugp}[UGP]{User Genre Profile}
\acrodef{agp}[AGP]{Age Genre Profile}
\acrodef{rp3beta}[\texttt{RP$^3\beta$}]{\texttt{RP$^3\beta$}}
\acrodef{children}[\texttt{child}]{\texttt{chil\-dren}}
\acrodef{mainstream}[\texttt{mainstream}]{\texttt{main\-stream users}}
\acrodef{nma}[\texttt{NMA}]{Non-Mainstream Adult}
\acrodef{le}[LE]{listening event}
\acrodef{bx}[BX]{Book Crossing}
\acrodef{ml}[ML]{MovieLens-1M}
\acrodef{jsd}[JSD]{Jensen-Shannon Divergence}
\acrodef{rp3}[\texttt{RP}$^3\beta$]{\texttt{RP}$^3\beta$}
\acrodef{reference}[reference]{reference framework}
\acrodef{refWork}[reference]{reference work}

\begin{document}

\title[Impacts of Mainstream-Driven Algorithms on Recommendations for Children Across Domains]{Impacts of Mainstream-Driven Algorithms on Recommendations for Children Across Domains: A Reproducibility Study}

\author{Robin Ungruh}
\email{R.Ungruh@tudelft.nl}
\orcid{0009-0004-4787-8897}
\affiliation{%
  \institution{Delft University of Technology}
  \city{Delft}
  \country{The Netherlands}
}

\author{Alejandro Bellog\'in}
\email{alejandro.bellogin@uam.es}
\orcid{0000-0001-6368-2510}
\affiliation{%
  \institution{Universidad Autónoma de Madrid}
  \city{Madrid}
  \country{Spain}
}

\author{Dominik Kowald}
\email{dkowald@know-center.at}
\orcid{0000-0003-3230-6234}
\affiliation{%
  \institution{Know Center Research GmbH \& TU Graz}
  \city{Graz}
  \country{Austria}
}

\author{Maria Soledad Pera}
\email{M.S.Pera@tudelft.nl}
\orcid{0000-0002-2008-9204}
\affiliation{%
  \institution{Delft University of Technology}
  \city{Delft}
  \country{The Netherlands}
}

\renewcommand{\shortauthors}{Ungruh, Bellog\'in, Kowald, and Pera}

\begin{abstract}
Children are often exposed to items curated by recommendation algorithms. Yet, research seldom considers children as a user group, and when it does, it is anchored on datasets where children are underrepresented, risking overlooking their interests, favoring those of the majority, i.e., mainstream users.
Recently, Ungruh et al. demonstrated that children's consumption patterns and preferences differ from those of mainstream users, resulting in inconsistent recommendation algorithm performance and behavior for this user group. 
These findings, however, are based on two datasets with a limited child user sample. We reproduce and replicate this study on a \textit{wider range of datasets} in the movie, music, and book \textit{domains}, uncovering interaction patterns and aspects of child-recommender interactions consistent across domains, as well as those specific to some user samples in the data. We also extend insights from the original study with \textit{popularity bias metrics}, given the interpretation of results from the original study. With this reproduction and extension, we uncover consumption patterns and differences between age groups stemming from intrinsic differences between children and others, and those unique to specific datasets or domains. 

\end{abstract}



\maketitle

\begingroup
\renewcommand\thefootnote{}\footnotetext{%
  \hspace{-1.5em}\raisebox{5pt}{%
    \begin{minipage}[t]{\columnwidth}
      \footnotesize
      © Robin Ungruh, Alejandro Bellog\'in, Dominik Kowald, and Maria Soledad Pera 2025. This is the author's version of the work. It is posted here for your personal use. Not for redistribution. The definitive Version of Record was accepted for publication in the \textit{19th ACM Conference on Recommender Systems (RecSys 2025)}, \url{https://doi.org/10.1145/3705328.3748160}
    \end{minipage}%
  }%
}
\endgroup

\section{Introduction}
Children, a heterogeneous population with unique preferences in media~\cite{milton2020don, ungruh2025impact, schedl2019online} and interaction patterns with online systems~\cite{viros2024can, konca2022digital}, are frequently exposed to decisions made by \ac{rs} on the various online platforms they use.
\ac{rs} research, however, rarely has them as the protagonist. Typically, design and evaluation of \ac{rs} are informed by data capturing user-system interactions from a broad user group, where children remain a minority. 
As the preferences of minorities tend to visibly deviate from the majority---a fact \ac{rs} may not capture~\cite{kheya2025unmasking,said2018coherence, mansoury2020investigating,eskandanian2019power}---and with children being vulnerable and in-development users~\cite{hargreaves2015and, valkenburg2001development}, explicit attention must be paid to their preferences and interaction patterns with \ac{rs}~\cite{ungruh2025impact, gomez2021evaluating, landoni2024good}.
As most knowledge about \ac{rs} stems from mainstream users~\cite{kowald2021support, kowald2022popularity, zhu2022fighting, li2021leave, ekstrand2018all}, concerns arise as to whether \acp{ra} can adequately accommodate the needs of an underrepresented group like children or if suggestions are inherently skewed toward mainstream actions. 

Recently, \citet{ungruh2025impact} introduced a \acl{reference} to study genre preference deviations between children and mainstream users and how \ac{rs} treat these user groups. They apply this framework to the \ac{ml}~\cite{harper2015movielens} and LFM-2b~\cite{schedl2022lfm} datasets in two experiments---one focused on age-related user preferences, and one on the dynamics between \ac{rs} and children.
Outcomes reveal \textit{deviating preferences} between children and mainstream users in the music and movie genres they consume, and that, while most of the \acp{ra} studied produce music recommendations that reasonably align with child preferences, for certain \acp{ra} the dominance of mainstream users in the training data \textit{skews} recommendations away from children's preferences. 

The at times conflicting findings, exacerbated by the limited datasets used---the main one no longer available---jeopardize result generalizability and call for additional analysis.
This prompts us to conduct a reproducibility study; elaborating on how children differ from the mainstream, and whether \acp{ra} account for emerging differences. Our motivation is rooted in three main pillars:

\noindent \textbf{1. Children as Non-Mainstream Users.} 
\citet{ungruh2025impact} highlight that across \ac{ml} and LFM-2b, children's consumption of items of different genres differs from those of mainstream users. As these outcomes are limited to two datasets and domains, it is unclear whether such deviations are unique to the datasets studied or represent a broader trend observable across datasets.

\noindent \textbf{2. Dominance of User Groups \& Deviating \ac{ra} Behavior.} 
Mainstream users' prominence in data used to train \acp{ra} affect how well these systems fare for children, i.e., the quality of the recommendations presented to this group \cite{ungruh2025impact}. 
As this is based on a snapshot from LFM-2b, this takeaway may be limited to the domain studied, particularly considering that music consumption has unique characteristics compared to other common domains in \ac{rs} research~\cite{schedl2021music}. Additionally, the analysis is restricted to a sample of users and a limited timeframe, raising critical concerns about the generalizability of findings. 
It also remains unclear why certain preferences are effectively captured while others are not. The \acl{refWork} offers interpretations of which interaction patterns may lead to better genre alignment between recommendations and preferences, other aspects are not studied, and assumptions require further validation.

\noindent \textbf{3. Reproducibility Concerns.} 
Thorough reproducibility 
is a challenging endeavour~\cite{bellogin2021improving, ferrari2021troubling,semmelrock2025reproducibility}. 
For the \acl{refWork}, this is further complicated by the unavailability of the LFM-2b dataset, making direct reproducibility impossible.
But even replicability, conducting the same experiments with different datasets, is not a straightforward process, considering the limited data sources available that include children, required data properties, and available metadata to conduct the study. \citet{ungruh2025impact} already recognize limitations in data processing and gathering for child-centered analyses; moreover, the \acl{reference} is tailored for a specific use case and datasets, giving cause to inquire about the feasibility of closely replicating the original setup with other datasets.

\noindent \textbf{This Work.} 
Driven by the aforementioned motivations, in this work, we (1) \textbf{reproduce} the \acl{refWork} to probe the results obtained by the original study, (2) \textbf{replicate} it by conducting the studied experiments with new datasets and a new domain to broaden insights and generalizability of findings from the \acl{refWork}, and (3) \textbf{extend} the analysis to 
explore additional facets of children's and mainstream users consumption patterns. 
In doing so, we broaden the scope of the original study by further exploring child and mainstream preferences within the context of \ac{rs} in three distinct domains: (1) \textit{movies} using \ac{ml}, (2) \textit{music} using MLHD~\cite{vigliensoni2017music}---a dataset that has received little attention by the \ac{rs} community, which becomes particularly relevant as an alternative to LFM-2b, allowing us to verify children's consumption trends in a different dataset in the music domain---, and (3) \textit{books} using \ac{bx}, a domain not considered by the original study. 

Although children are not a uniform user group, we often refer to children as \textit{one} group to create a foundational understanding of the interplay between children and \ac{rs}. This simplification aids understanding of children's role as non-mainstream users in a landscape increasingly shaped by ubiquitous \acp{ra}. Children are a vulnerable user group. As such, ethical considerations and deliberation are important. Thus, we only utilize publicly available datasets where users voluntarily self-declared age information, i.e., information about users was crawled and aggregated in the used datasets.

\noindent \textbf{Contributions.} This work expands the \acl{reference} for an extended picture about children and mainstream users. With our multi-domain study, we probe trends consistent with the \acl{refWork} as well as those deviating.
Generalization to other datasets advances knowledge on how \acp{ra} fare for non-mainstream user groups more broadly; auditing their ability to serve \textit{each} user well.

\noindent \textbf{Reproducibility.} 
We provide code to reproduce our experiments (\textcolor{blue!75}{\faGithub}~\url{https://github.com/rUngruh/2025_RecSys_Reproducibility}). We publish the used sample of MLHD, enabling easier filtering for future studies (\url{https://zenodo.org/records/15394228}).

\section{Reference Work}\label{sec:replicated_work}

The \acl{refWork}~\cite{ungruh2025impact} provides a \textbf{\acl{reference}} along with associated code to enable reproducibility. This framework has the following key components: a \textbf{dataset} in a given domain, a classification of users into distinct \textbf{user groups}---including a method to identify one as \textbf{mainstream}---and a way to quantify \textbf{preference} alignment, with a focus on the alignment between user preferences and recommendations generated by a set of \textbf{\acp{ra}}. 
The framework is applied on two experiments: (1) the \textit{Preference Deviation Exploration}, which determines the degree to which groups differ from the mainstream, and (2) the \textit{RS experiment}, which examines the impact of mainstream users' presence in the source data on the quality of recommendations for an underrepresented group.

The \acl{reference} compared the genres of consumed and recommended items of users in different age groups, focused on the user group of children. 
This was based on two datasets, (1) \ac{ml}, which captures movie ratings, and (2) LFM-2b, which explores music listening events. By grouping users based on their age, the authors categorize \acl{mainstream} as users belonging to the most common age group---young adults who contribute the majority of data---\acl{children} are younger minors, while \acfp{nma} are adults older than the \acs{mainstream}. In both datasets, \acl{children} are only responsible for a minority of the user-item interactions ($2.83\%$ and $7.07\%$, respectively). To investigate differences in preferences, as well as the degree to which these preferences are captured by common \acp{ra}, the study assesses user preferences by the genres of items previously consumed by a user to measure genre alignment, a concept closely related to \textit{miscalibration}~\cite{steck2018calibrated}.

The \textit{Preference Deviation Exploration} surveys the differences between age groups, in particular between \acl{children} and \acl{mainstream}, across both datasets. Aggregating the preferences of users belonging to a certain age group enabled assessment of how much users within one age group deviate from each other, but also analysis of the degree to which age groups differ from each other. Outcomes showed that, regardless of the dataset, genre consumption differs between age groups, with \acl{children}'s genre preferences differing markedly from those of \acl{mainstream}.

The \textit{RS experiment} considers the top-$50$ recommendations created by varied \acp{ra} for a sample from LFM-2b. To gauge whether these recommendations align with the preferences of users of different ages, the analysis---by age group---leverages classical performance metrics and the alignment of genres between users' previous consumption and recommendations. To probe the impact of \acl{mainstream} in the train data on the underrepresented group, a two-step evaluation is adopted: \acp{ra} are first trained on a \texttt{General Set} that includes the interactions of users of varying ages (where \acl{mainstream} dominate the data) and once on a \texttt{Child Set} that only includes interactions of \acl{children}. As the \acp{ra} do not have any \acs{mainstream} data available in this latter setup, performance in the two recommendation scenarios can be compared to assess the impact of \acl{mainstream} on recommendations for \acl{children}.

\section{Reproducibility, Replicability, and Extension}

Our work aims to both address constraints of the \acl{refWork} and contextualize replicated results with additional explanatory factors. As the \acl{refWork} published associated code, we can follow the \acl{reference}, only making adaptations to improve the clarity of the code and facilitate integration of new datasets. An overview of our reproducibility efforts can be seen in \cref{tab:reproducibility_approach}.

\begin{table}[h]
\caption{Overview of reproducibility ($repr$), replicability ($repl$), and  extension ($ext$) efforts across datasets.}
\label{tab:reproducibility_approach}
\footnotesize
\begin{tabular}{llcccc}
    \toprule
    \multirow{2}{*}{Domain} & \multirow{2}{*}{Dataset} & \multicolumn{1}{c}{New} & \multicolumn{1}{c}{New} & \multicolumn{1}{c}{Pref. Dev.} & \multicolumn{1}{c}{RS} \\
     &  & \multicolumn{1}{c}{Domain} & \multicolumn{1}{c}{Dataset} & \multicolumn{1}{c}{Exploration} & \multicolumn{1}{c}{Experiment} \\
    \hline \hline
    Movies & \acs{ml} & No & No & $repr$ + $ext$ & $repl$ + $ext$ \\
    Music & MLHD & No & Yes & $repl$ + $ext$ & $repl$ + $ext$ \\
    Books & \acs{bx} & Yes & Yes & $repl$ + $ext$ & $repl$ + $ext$ \\
    \bottomrule
\end{tabular}
\end{table}

We \textit{reproduce} the \textit{Preference Deviation Exploration} experiment with \ac{ml} to probe the original analysis and verify if our study leads to the same results. Recall that we exclude LFM-2b as it is no longer available. Further, we \textit{replicate} this experiment on two datasets new to this study: MLHD~\cite{vigliensoni2017music} as an alternative to LFM-2b to explore users' interactions with songs, and \acf{bx}~\cite{ziegler2005improving}, which tracks book interactions---a domain not considered by the \acl{refWork}. Each dataset was chosen as it includes demographic information about users and can be annotated with item genres, which are leveraged to determine preferences. Together, this enables juxtaposing deviating preferences in age groups across domains.

Inconsistencies in findings of the \textit{RS experiment} and insights limited to a restricted period in LFM-2b prompt replication of this experiment on \ac{ml}, MLHD, and \ac{bx}. This allows probing of the stability of outcomes and the effect of \acl{mainstream} on recommendations for \acl{children} across various datasets and domains.

Genre consumption behaviour is \textit{one} aspect that may trigger differences across user groups; the \acl{refWork} indicates that other consumption characteristics could be distinguishing factors for varying \ac{ra} behavior between \acl{children} and \acl{mainstream}. 
As popularity oftentimes affects recommendation quality~\cite{kowald2022popularity, abdollahpouri2021user, kowald2020unfairness, abdollahpouri2019unfairness}, we explore popularity as a potential factor contributing to deviating \ac{ra} behavior. To do so, we \textit{extend} the \acl{reference} with a \textit{popularity extension} in both experiments. For the \textit{Preference Deviation Exploration}, we analyze the popularity of items consumed by users of different ages to assess the degree to which users consume items that are (1) overall popular, or (2) popular among users of the same age range. In the \textit{RS experiment}, we gauge if \acp{ra} amplify popularity in recommendations compared to users' profiles.

\section{Experimental Setup}
We discuss the datasets and set up. We explicitly highlight adaptations made to the \acl{refWork} required to facilitate our study.

\begin{figure*}[t]
    \centering
    \hfill
    \begin{subfigure}[t]{0.32\textwidth}
        \centering
        \includegraphics[width=\textwidth]{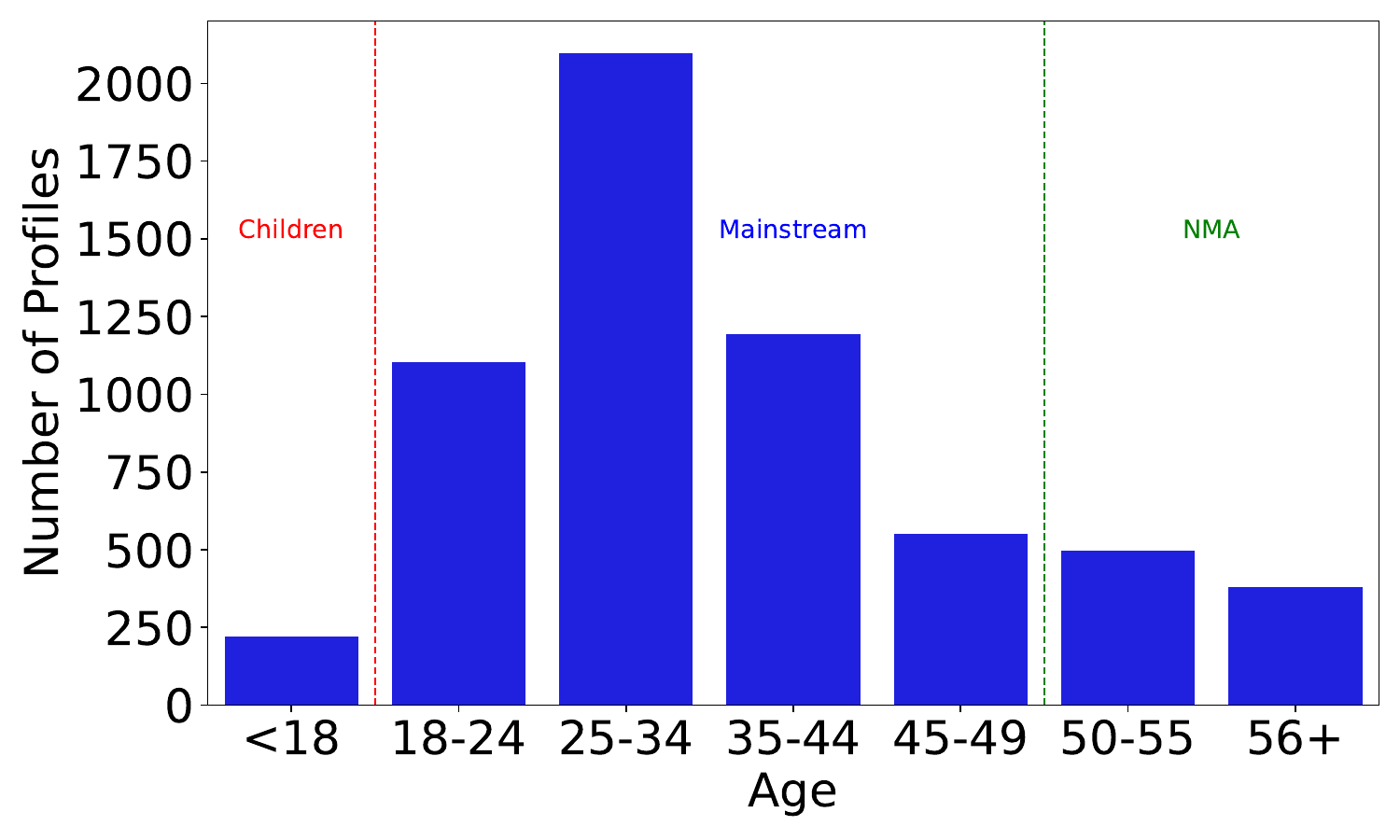}
        \caption{Users per age in \ac{ml}.}
        \label{fig:setup:num_profiles:ML}
    \end{subfigure}
    \hfill
    \begin{subfigure}[t]{0.32\textwidth}
        \centering
        \includegraphics[width=\textwidth]{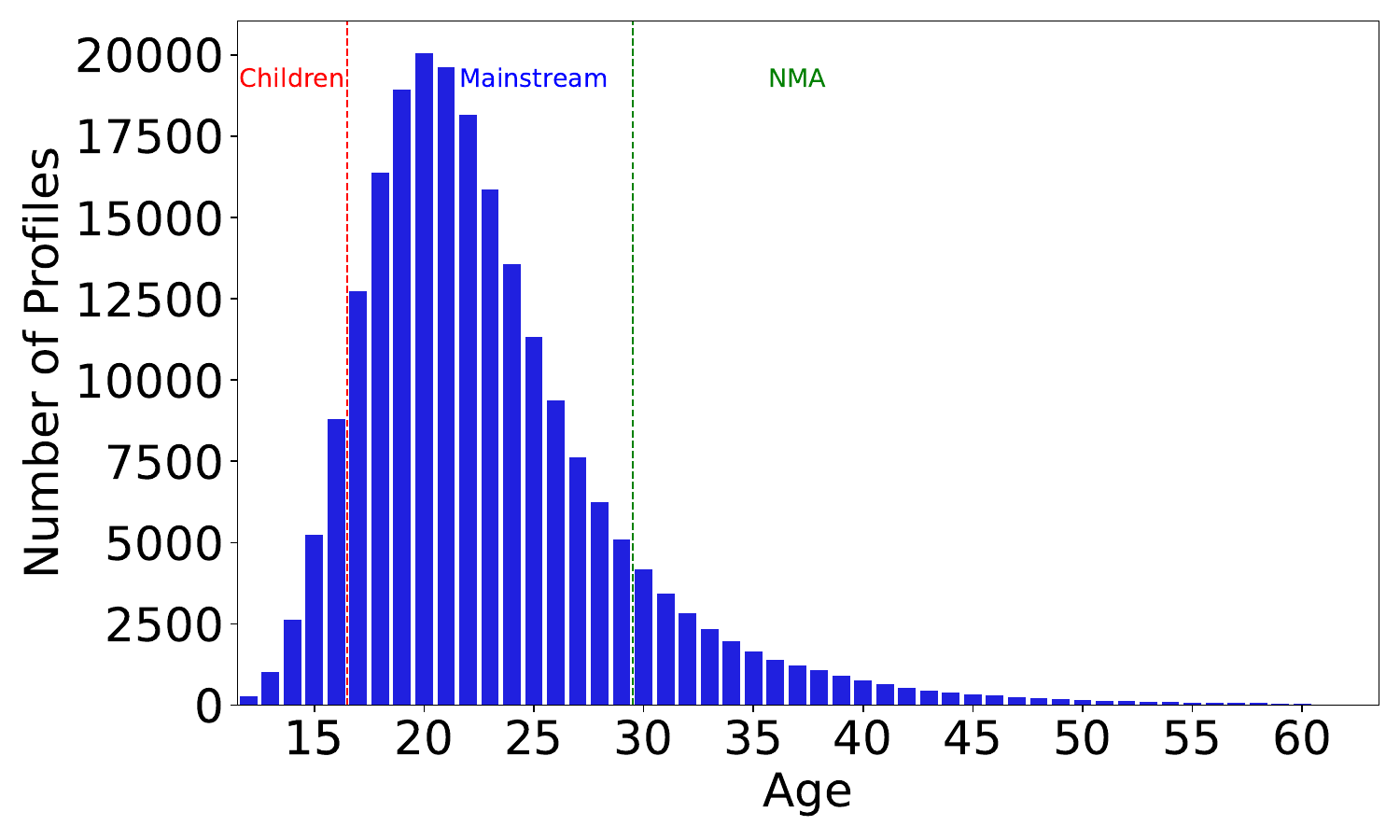}
        \caption{Users per age in MLHD.}
        \label{fig:setup:num_profiles:MLHD}
    \end{subfigure}
    \hfill
    \begin{subfigure}[t]{0.32\textwidth}
        \centering
        \includegraphics[width=\textwidth]{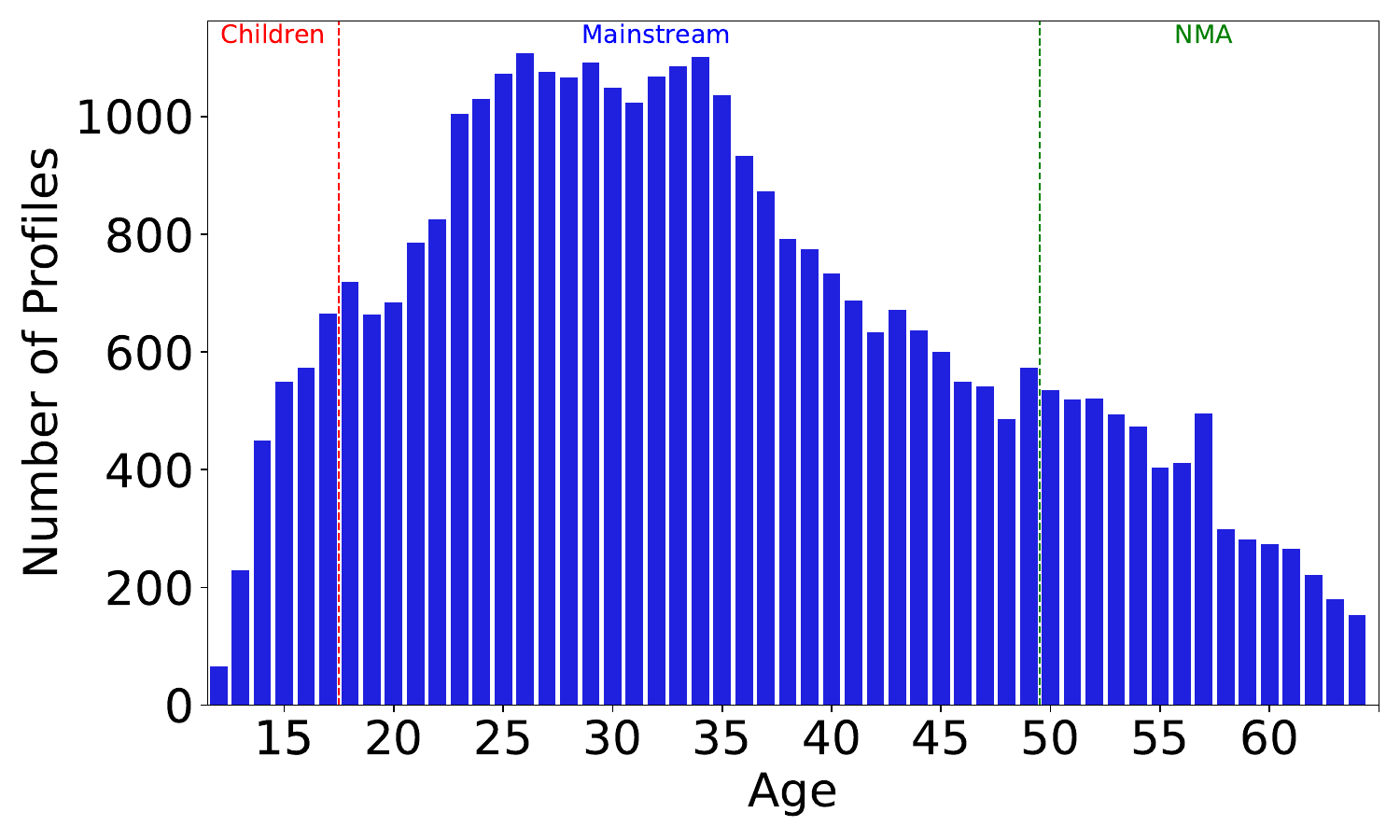}
        \caption{Users per age in \ac{bx}.}
        \label{fig:setup:num_profiles:BX}
    \end{subfigure}
    \hfill
    \caption{Size comparisons of the datasets.}
    \label{fig:setup:num_profiles}
\end{figure*}

\noindent \textbf{Datasets.} 
As in \cite{ungruh2025impact}, we preprocess the \textbf{datasets} as described below and extend them with genre information; we only consider users aged $12$ to $65$, removing interactions with items with no assigned genre and users without valid age information. An overview of the resulting datasets is presented in \cref{tab:methodology:datasets} and \cref{fig:setup:num_profiles}.

\begin{table}[t]
    \centering
    \caption{Dataset description, including information on the number of \acs{children} and \acl{mainstream} (MS) in the datasets.}
    \label{tab:methodology:datasets}
    \footnotesize
    \begin{tabular}{lrrrrr}
            \toprule
        Dataset & \# Users  & \% \texttt{Child} & \% \texttt{MS} & \# Items & \# Interactions \\
        \hline \hline
         \ac{ml} & 6,040 & 3.68 & 81.82 & 3,706 & 1,000,209 \\
         \ac{bx} & 35,029 & 7.23 & 77.00 & 80,785 & 396,460 \\
         MLHD & 44,349 & 8.21 & 79.89 & 1,918,414 & 1,055,574,094 \\
        \bottomrule
    \end{tabular}

\end{table}

\textbf{\acf{ml}}~\cite{harper2015movielens} 
includes ratings for movies, where each movie is annotated with at least one of $18$ genres that we assign equal weighs; users are assigned to one of seven age groups. As per the \acl{refWork}, we treat users with the label `Under 18' as \acl{children}, users aged 18 to 49 as \acl{mainstream}, and older users as \ac{nma}. Only $2.83\%$ of ratings can be attributed to \acl{children} and $84.60\%$ to \acl{mainstream}.

\textbf{MLHD}~\cite{vigliensoni2017music} includes $27$ billion 
interactions with tracks, gathered from Last.fm (\url{https://www.last.fm/}). We utilize MLHD+\footnote{\url{https://musicbrainz.org/doc/MLHD+}}, an improved version of the dataset that allows simple matching to information provided by MusicBrainz. 
We follow common practice regarding sampling large datasets for experimentation~\cite{lesota2021analyzing, lesota2024oh}, and select a user sample of the dataset. 
To capture users with consistent interactions over an extended period, we select those whose first recorded interaction occurred in or before $2009$ and the last in or after $2013$. This five-year window captures the period during which most interactions occurred, as well as the most recent span where user activity was consistently recorded. In line with~\cite{kowald2021support}, we exclude users with an unusually high number of interactions, i.e., those whose number of recorded interactions exceeds the mean by more than two standard deviations in the dataset.
From the remaining users, we randomly sample $45{,}000$ users---a comparable number of users to those considered in the \acl{refWork} for the music-related dataset---while preserving the age distribution of the unfiltered dataset to maintain its original demographic structure.

Songs in MLHD are not linked to a genre. Hence, we annotate each song based on artist genres from Allmusic, as done for LFM-2b\footnote{Allmusic updated their genres (\url{https://www.allmusic.com/genres}). For comparability to the original study, we use the genres used in~\cite{schedl2017large}.}. 
For this, we extract artist genres using the MusicBrainz API\footnote{\url{https://musicbrainz.org/doc/MusicBrainz_API\#Lookups}} and match these fine-grained genres with Allmusic genres, annotating artists with at least one of the $20$ genres. In line with the \acl{refWork}, we assume that the genre distribution of each artist extends to their tracks and annotate each song with equally weighted genres.

As the age information is ``the age returned by the system at the moment of the data collection (i.e., circa 2013 and 2014)''\footnote{\scriptsize{\url{https://ddmal.music.mcgill.ca/research/The_Music_Listening_Histories_Dataset_(MLHD)/}}}, we assume that each user turned the reported age on January $1$st, $2014$. This enables us to assess each user's age for each interaction. For age groupings, we follow those used in LFM-2b in the \acl{refWork}, as MLHD exhibits a similar age distribution: 
\acl{mainstream} are users aged $17$ to $29$, as this age range accounts for the vast majority of interactions in the dataset. Although $17$-year-olds are legally considered children per UNICEF's definition~\cite{UNICEF}, they cannot be treated as a minority in this context due to their high representation in the data. Consequently, we categorize users \textit{under $17$} as \acl{children}. Users older than $29$ are referred to as \acp{nma}.
Most interactions in the dataset are from \acl{mainstream} ($78.89\%$), while only a $10.80\%$ of interactions can be attributed to \acl{children}.

\textbf{\acf{bx}}\footnote{\url{https://www.kaggle.com/datasets/syedjaferk/book-crossing-dataset}} \cite{ziegler2005improving} contains over $1$ million user-book interactions. 
As books in \ac{bx} lack genre annotations, we combine \ac{bx} with the Goodreads dataset\footnote{\url{https://cseweb.ucsd.edu/~jmcauley/datasets/goodreads.html}} \cite{wan2018item, wan2019fine}, which links books with at least one out of $8$ genres. 
Books are assigned different ISBNs depending on their editions; thus, we turn to the \texttt{bookdata tool (3.0)} \cite{ekstrand2018exploring} to obtain ISBN variations for each book in Goodreads and \ac{bx}, which we use for merging purposes. We assign equal weighting to each genre a book is annotated with. 
Age distributions are akin to those on \ac{ml}. Thus, we also treat users with age annotated $<18$ as \acl{children}, users aged $18$ to $49$ as \acs{mainstream}, and older users as \acsp{nma}. On \ac{bx}, only $2.88\%$ of interactions can be attributed to \acl{children} and $81.29\%$ to \acl{mainstream}.


\noindent \textbf{Preference Deviation Exploration.}  
We create \textit{user profiles} that capture all items a user consumed throughout a year of their lives. As \ac{bx} and \ac{ml} do not include timestamps, we assume that each interaction is associated with the reported age of the user, resulting in one user profile per user; for MLHD, a yearly user profile is created for each year in which a user interacted with tracks.

Genre preferences within user profiles are analyzed using \textbf{User Genre Profiles} ($UGP$s), which represent the mean frequency of each genre in a user’s profile, accounting for genre weightings~\cite{schedl2017large}. $UGP$s are used as a proxy to assess differences in preferences by accounting for varying consumption patterns. Multiple consumptions of the same item are considered to account for repeated interactions. 
To gain insights into broader consumption patterns of age groups, \textbf{Age Genre Profiles} $AGP_{age}$ represent the average genre consumption of users within a particular age bucket, denoted with $age$. 
We quantify differences in genre consumption with: 
\begin{itemize}[nosep, wide, left=0pt .. \parindent]
    \item \textbf{In-group Deviation} ($IGD_{age}$) assesses how much profiles within an age group deviate from each other by measuring the average \ac{jsd} between an $AGP_{age}$ and each $UGP$ of users of the same $age$. This metric is analogous to a standard deviation as it measures the average distance of genre distributions of user profiles to the average distribution across all users in the respective age group.
    \item \textbf{Age Preference Deviation} ($APD_{age1, age2}$) measures the \ac{jsd} between two $AGP$s.
\end{itemize}
The \acl{refWork} compared significant differences between the genre consumption of \acl{children} and \acl{mainstream} using a MANOVA and matching post-hoc tests. This approach relies on assumptions such as multivariate normality and homogeneity of variance, which may not be fully met in the case of normalized genre proportions. To better accommodate the nature of the data, we adopt a non-parametric approach and use Kruskal–Wallis tests to assess overall group differences per genre, followed by pairwise Mann–Whitney U tests with multiple testing correction ($p<0.01$). 

\noindent \textbf{Popularity Extension.} To study the impact of popularity bias between users, we extend the \acl{reference} with:
\begin{itemize}[nosep, wide, left=0pt .. \parindent]
    \item \textbf{\# Interactions:} The number of items a user has consumed.
    \item \textbf{Profile Size:} The number of distinct items in a user profile \cite{kowald2020unfairness}.
    \item \textbf{Profile Popularity:} The average popularity of items in a user's profile, where an item's popularity is defined as the number of users who interacted with it, normalized by the most interacted-with item. This captures a user’s tendency to consume items that are broadly popular across the population.
    \item \textbf{Profile Age-Popularity:} The average age-specific popularity of items (the normalized number of same-age users who interacted with each item) in a profile. This captures a user's preferences for items popular among peers in their age group.
\end{itemize}

\noindent \textbf{RS Experiment.} 
To prepare the data for this experiment, the \acl{refWork} applies common preprocessing steps associated with music-\ac{rs} explorations: Restricting data to a specific timeframe, removing one-time listens to avoid spurious interactions~\cite{lesota2021analyzing, melchiorre2021investigating}, binarizing data by only including the first listening event~\cite{anelli2022top}, and applying $k$-core filtering to reduce sparsity in the data~\cite{anelli2022top}. Further, a temporal global split is used~\cite{campos2014time, kowald2013forgetting, ji2023critical} and users who lack items in any of the splits are removed. The training set resulting from this is the \texttt{General Set}; a subset only including \acs{children} interactions forms the \texttt{Child Set}.
We probe the \textbf{top-$50$ recommendations} created by two unpersonalized baselines---\texttt{Random} and \texttt{MostPop}---and two personalized \acp{ra}---\acl{rp3} \cite{paudel2016updatable} and \texttt{iALS} \cite{hu2008collaborative}.

To assess alignment between recommendations and user preferences, users' genre consumptions are captured by $UGP$s as defined previously; here, however, only interactions from the training sets are considered as these are the ones available to the \ac{ra}. $AGP$s are computed based on these $UGP$s. 
In line with $UGP$s, \textbf{Recommendation Genre Profiles} ($RGP$s) model the average genre distribution of items recommended to a user.
To gauge \ac{ra} performance, traditional performance metrics $nDCG$, $MRR$, $MAP$~\cite{tamm2021quality, ferrari2021troubling, gunawardana2022evaluating} are used in addition to \textbf{Genre Miscalibration} \cite{steck2018calibrated}---computed as 
the \ac{jsd} between a $UGP$ and the respective $RGP$---is computed. 

For the \textbf{popularity extension}, we add \textbf{Popularity Lift} ($PL$): The normalized difference between the Profile Popularity and the average popularity of items in recommendations for a user~\cite{kowald2020unfairness, abdollahpouri2019unfairness}.

For replication across datasets with different sizes and available metadata, we carefully design our setup to ensure comparability 
with the \acl{refWork} while accounting for the datasets' unique properties. As the datasets stem from widely different domains, most preprocessing steps are not universally applicable; hence, we discuss details and reasoning for deviations from the original setup.

We can follow the \acl{reference} closely for the \textit{RS experiment} using MLHD. However, due to the size of MLHD, we use a representative subset for the recommendation experiments---a common approach in music recommendation experiments~\cite{kowald2020unfairness}. We first create a subset of $13{,}000$ users who interacted with at least $5$ tracks while ensuring that we retain the original age distribution. Then, we restrict our data to the same months as in the \acl{refWork} (June to October) to avoid differences through seasonal effects. We set the year $2009$ for our exploration due to a more reliable number of young users within this year. 
We retain other preprocessing steps: We only consider user-song interactions where the user has listened to a song at least twice, and we binarize ratings. 
We remove items with fewer than $10$ and users with fewer than $5$ interactions. We split by using June to August for training, September for validation, and October for testing. Users lacking items in any of the splits are removed, resulting in a set of $10{,}325$ users with $97{,}322$ items. 
The \texttt{Child Set} includes $1{,}878$ users and $81{,}674$ items.

As interactions with movies and books differ markedly from interactions with music (particularly in the number of consumed items), we adapt our processing steps for \ac{ml} and \ac{bx}.
For \ac{ml}, we follow common practices~\cite{anelli2022top} of binarizing the data by only treating ratings $>3$ as positive signals, while we keep all interactions for \ac{bx} as positive signals, as it includes explicit as well as implicit interactions.
To reduce sparsity, we follow~\cite{anelli2022top} by applying iterative $k$-core with $k=10$ for \ac{ml}; as users tend to interact with fewer items on \ac{bx}, we set $k=5$ here.
For \ac{bx} (due to the lack of timestamps) and \ac{ml} (due to the short available timeframe~\cite{harper2015movielens}), a global temporal split is not applicable. Thus, we split users' interactions into $60\%$ training, $20\%$ validation, and $20\%$ test data. This results in a set of $5{,}949$ users with $2{,}810$ items for \ac{ml}, with a \texttt{Child Set} of $218$ users and $1{,}802$ items. The filtered set of \ac{bx} includes $6{,}950$ users and $16{,}477$ items, and a \texttt{Child Set} of $266$ users and $2{,}196$ items.

\noindent \textbf{RS Explorations \& Hyperparamenter Tuning.} 
We use the Elliot framework for RS explorations and tune the hyperparameters following the original study and~\cite{anelli2022top}. Besides aforementioned metrics, the \acl{refWork} explored additional metrics to gauge the deviation of $UGP$s and $RGP$s to \acs{mainstream} and \acs{children} $AGP$s. We bypass reporting these results as our focus is on \textit{direct comparisons} between age groups. 
However, we provide results for these metrics in our Git repository for transparency and completeness.

\section{Results}
Here, we lay out the outcomes from both experiments.

\begin{figure*}[]
    \centering
    \begin{subfigure}[t]{0.32\linewidth}
        \centering
        \includegraphics[width=\textwidth]{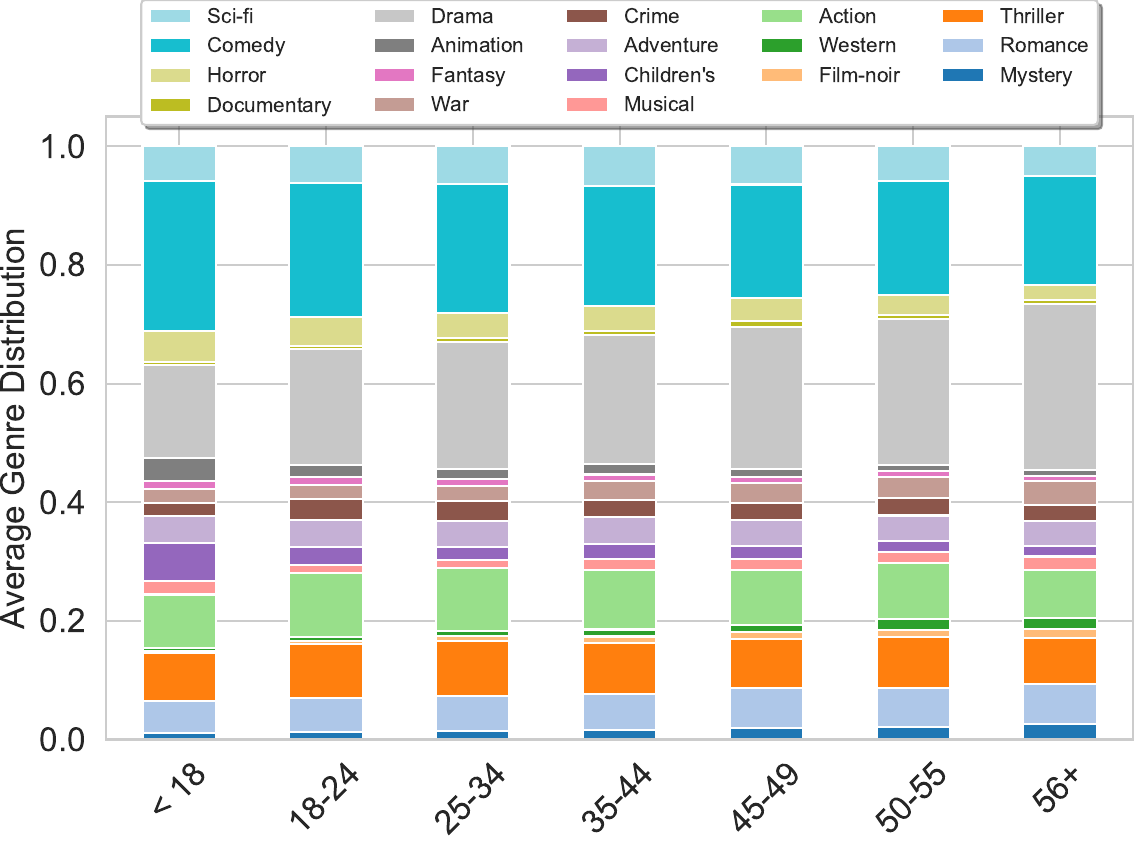}
        \caption{$AGP$s on \ac{ml}}
        \label{fig:results:experiment1:ml:genredistribution}
    \end{subfigure}
    \hfill
    \begin{subfigure}[t]{0.32\linewidth}
        \centering
        \includegraphics[width=\textwidth]{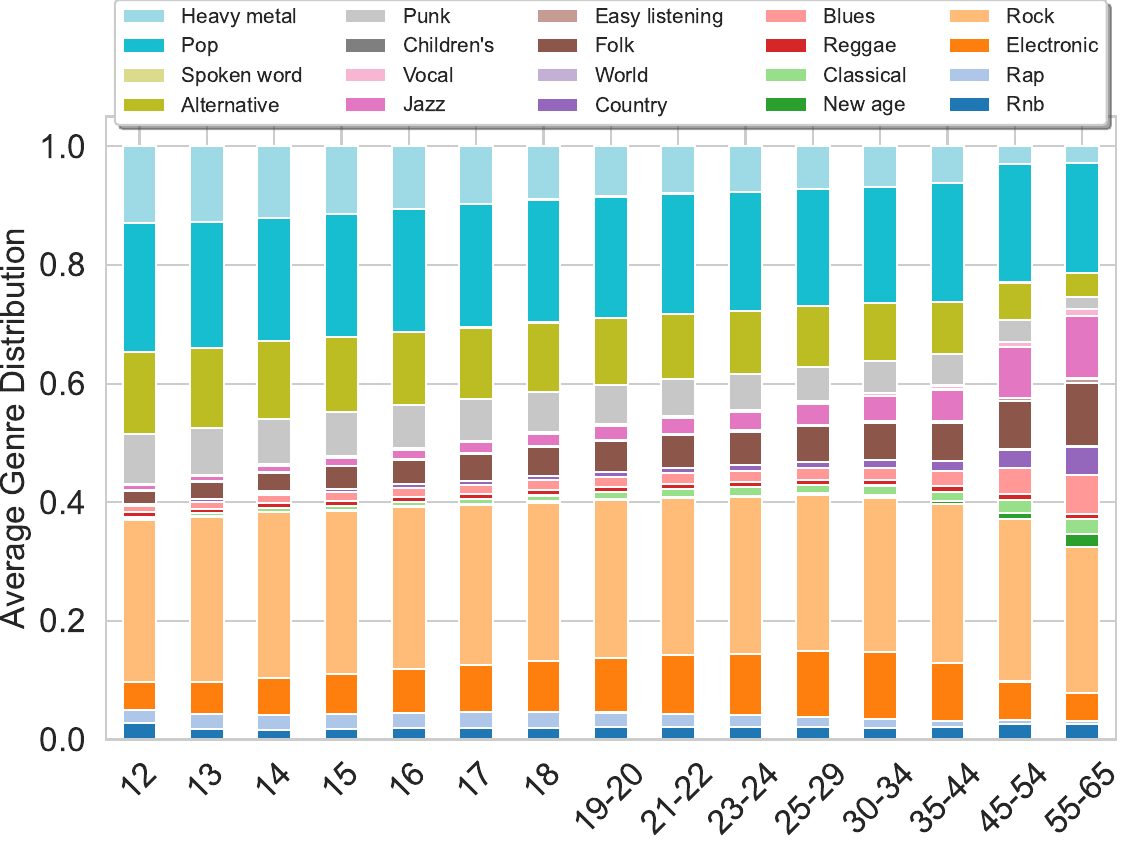}
        \caption{$AGP$s on MLHD}
        \label{fig:results:experiment1:mlhd:genredistribution}
    \end{subfigure}
    \hfill
    \begin{subfigure}[t]{0.32\linewidth}
        \centering
        \includegraphics[width=\textwidth]{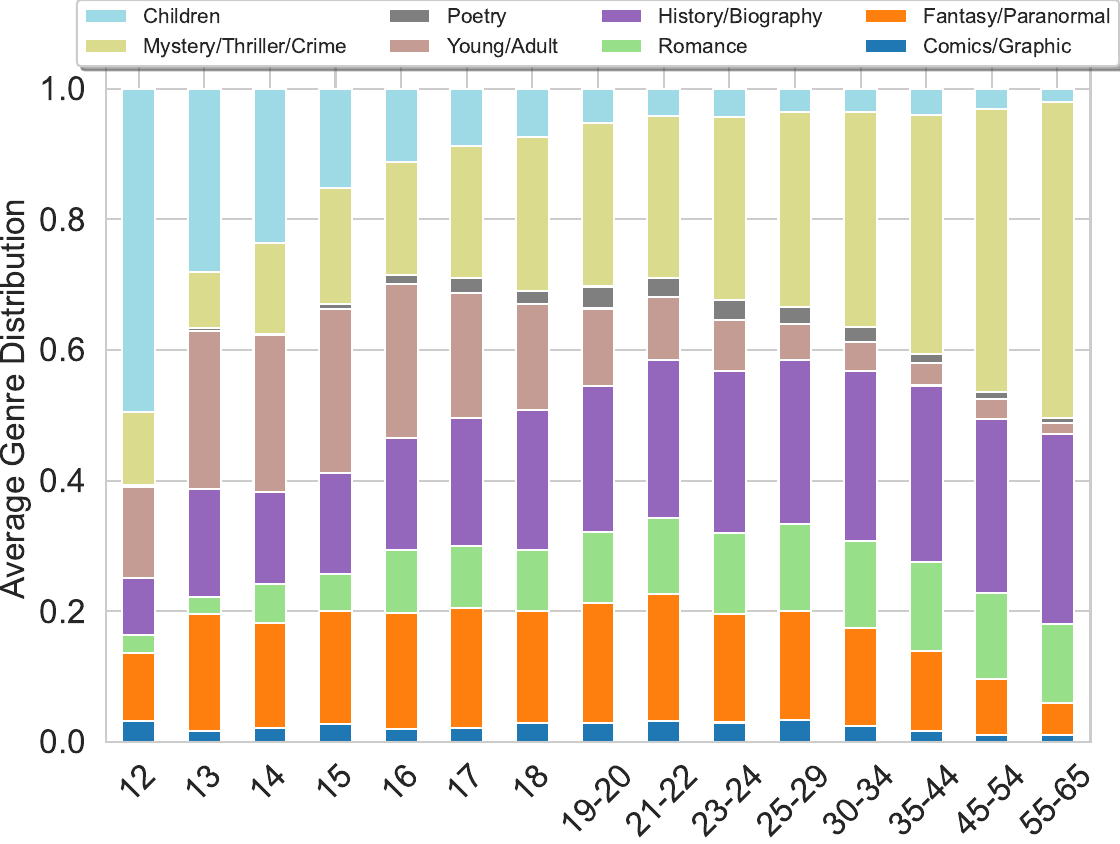}
        \caption{$AGP$s on \ac{bx}}
        \label{fig:results:experiment1:bx:genredistribution}
    \end{subfigure}    
    \begin{subfigure}[t]{0.275\linewidth}
        \centering
        \includegraphics[width=\textwidth]{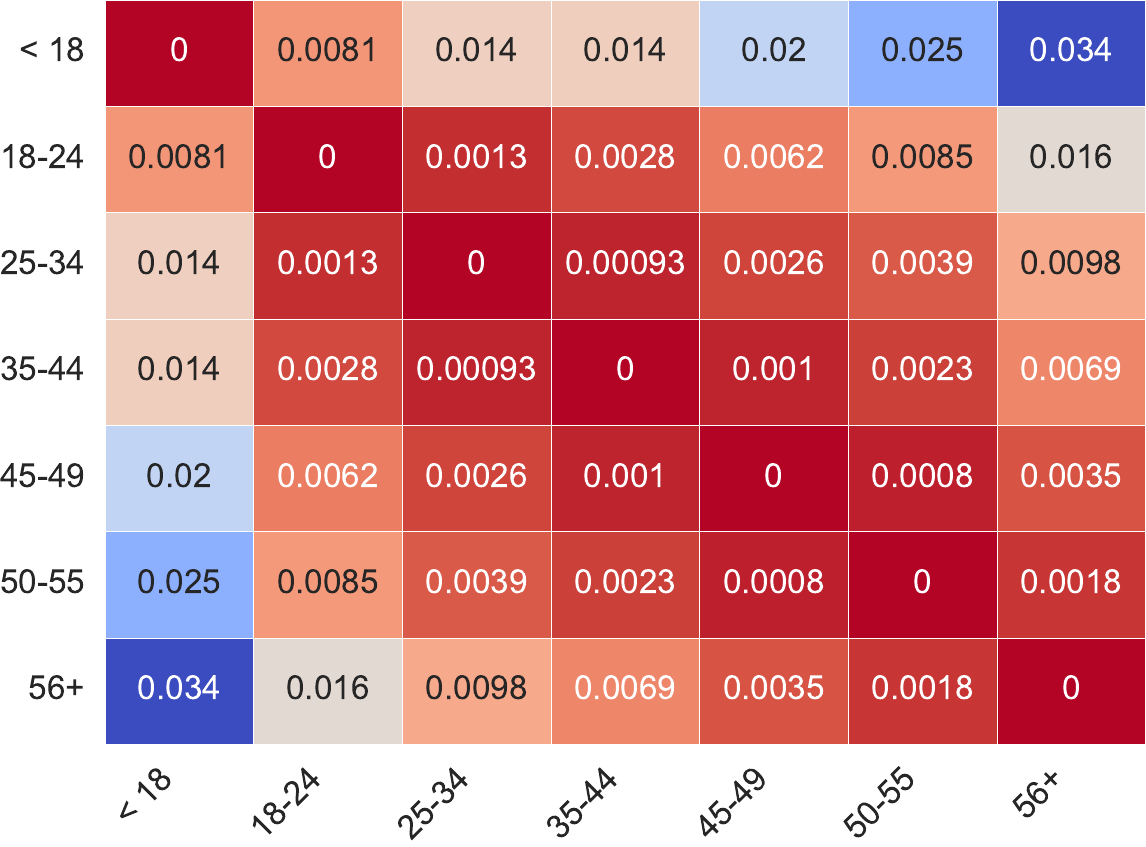}
        \caption{$APD$s on \ac{ml}}
        \label{fig:results:experiment1:ml:apd}
    \end{subfigure}
    \hfill
    \begin{subfigure}[t]{0.275\linewidth}
        \centering
        \includegraphics[width=\textwidth]{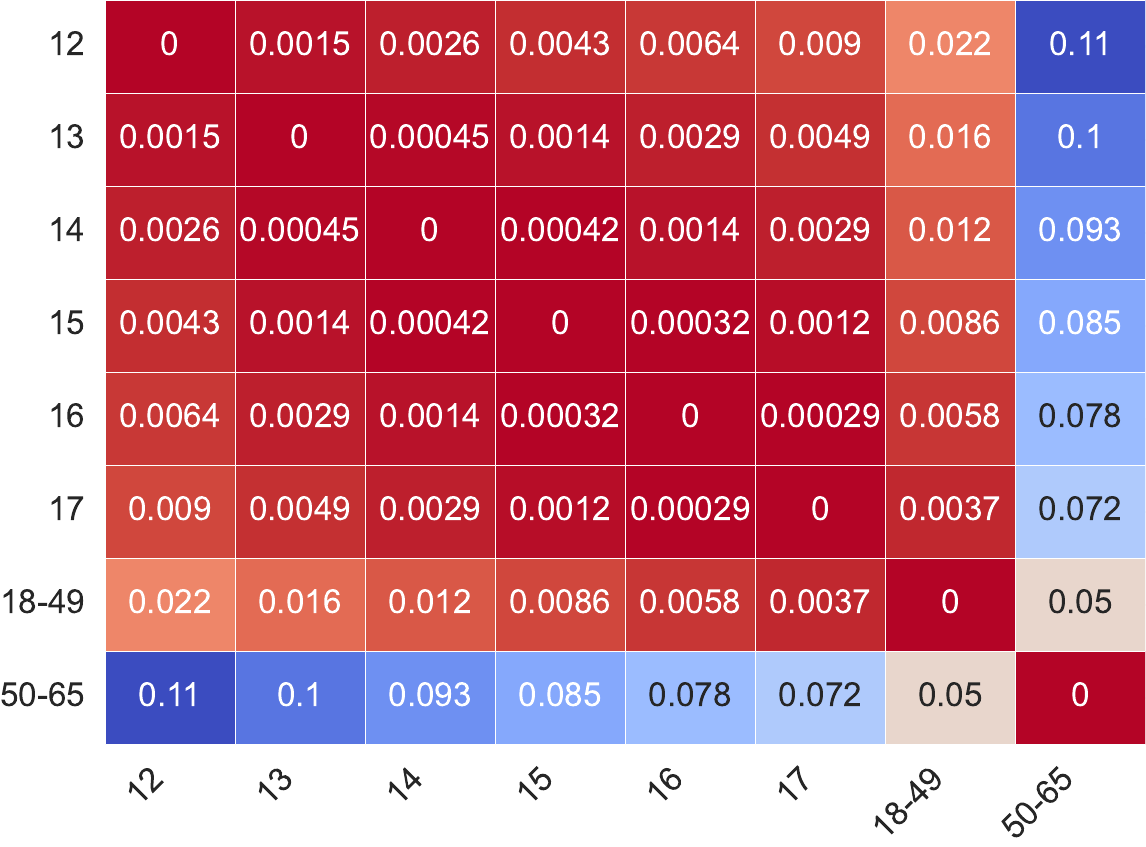}
        \caption{$APD$s on MLHD}
        \label{fig:results:experiment1:mlhd:agp}
    \end{subfigure}
    \hfill
    \begin{subfigure}[t]{0.275\linewidth}
        \centering
        \includegraphics[width=\textwidth]{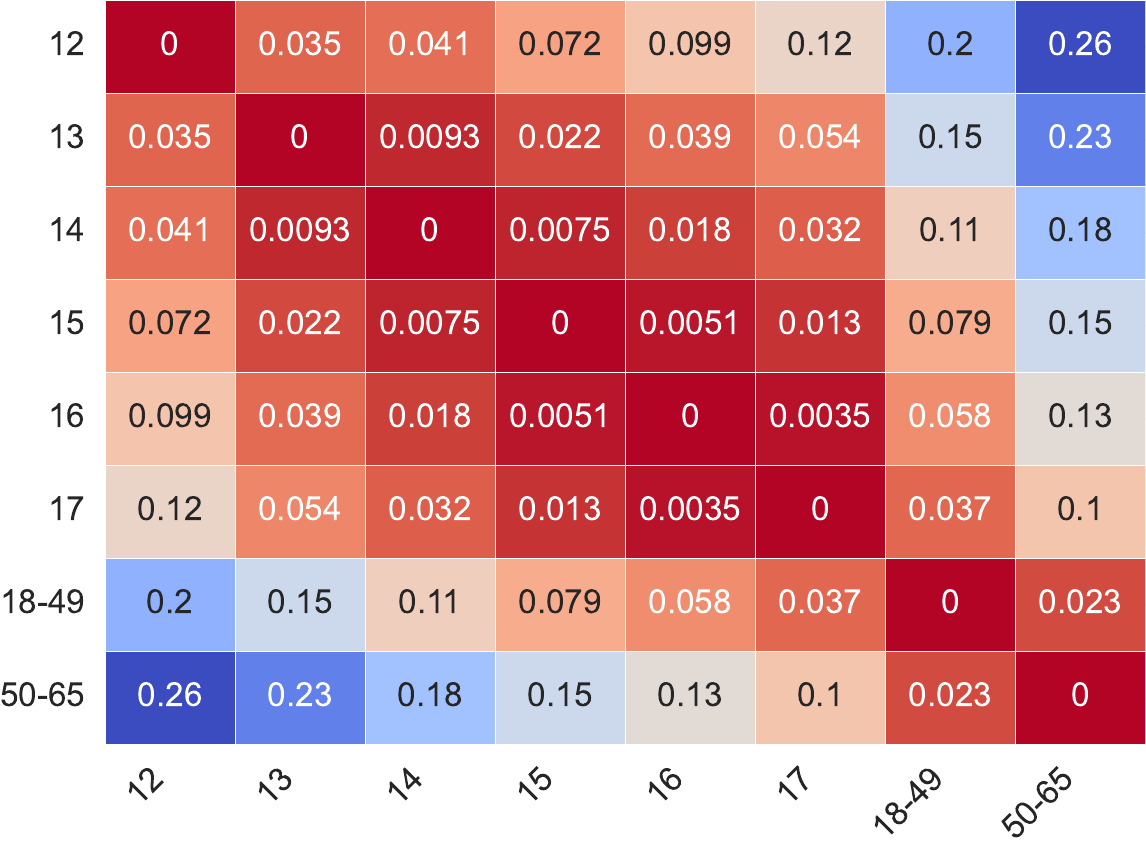}
        \caption{$APD$s on \ac{bx}}
        \label{fig:results:experiment1:bx:apd}
    \end{subfigure}
     \begin{subfigure}[t]{0.275\linewidth}
        \centering
         \includegraphics[width=\textwidth]{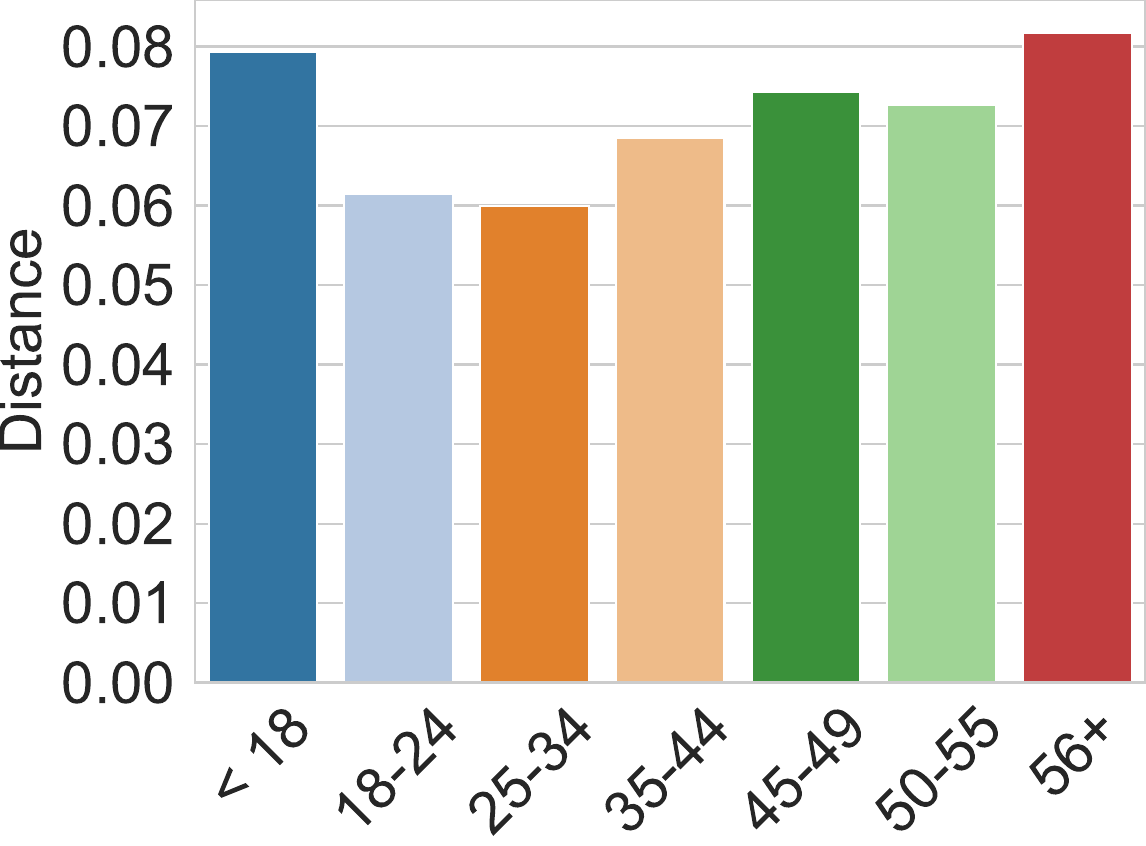}
        \caption{$IGD$ on \acs{ml}}
        \label{fig:results:experiment1:ml:igd}
    \end{subfigure}
    \hfill
    \begin{subfigure}[t]{0.275\linewidth}
        \centering
 \includegraphics[width=\textwidth]{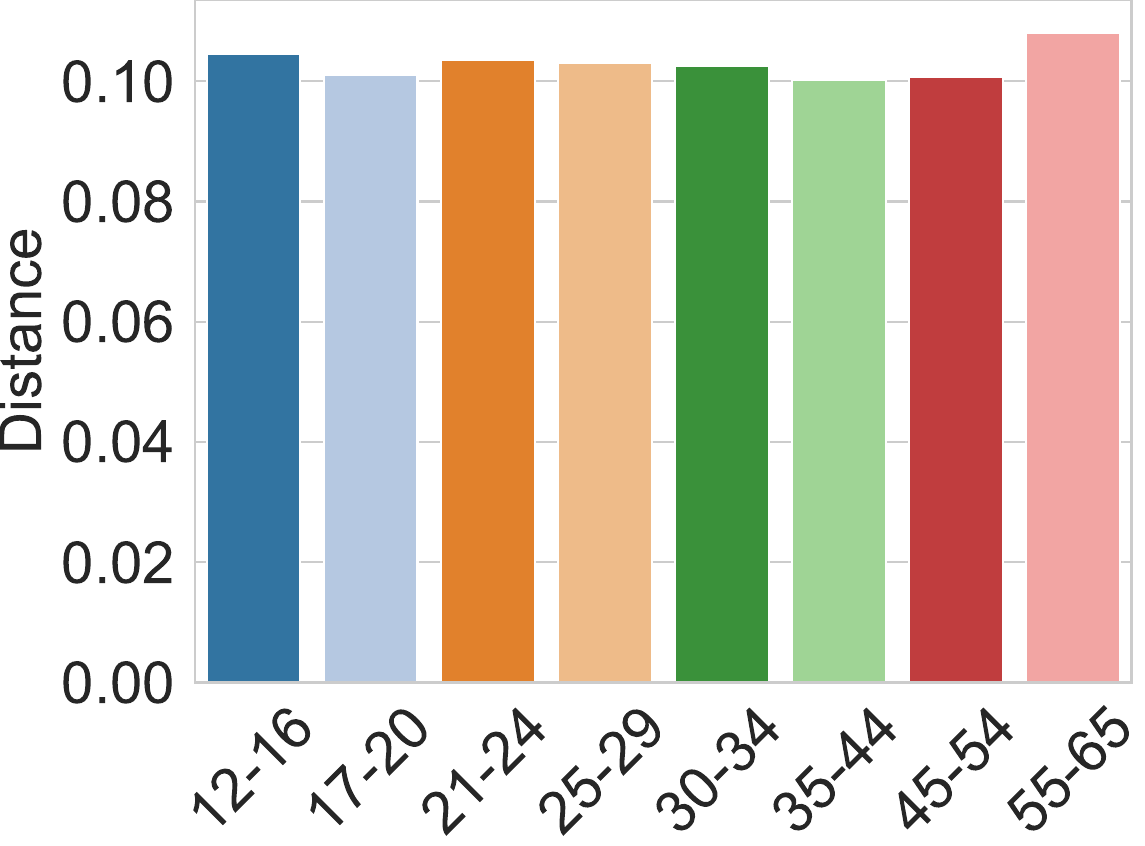}
        \caption{$IGD$ on MLHD}
        \label{fig:results:experiment1:mlhd:igd}
    \end{subfigure}
    \hfill
    \begin{subfigure}[t]{0.275\linewidth}
        \centering
         \includegraphics[width=\textwidth]{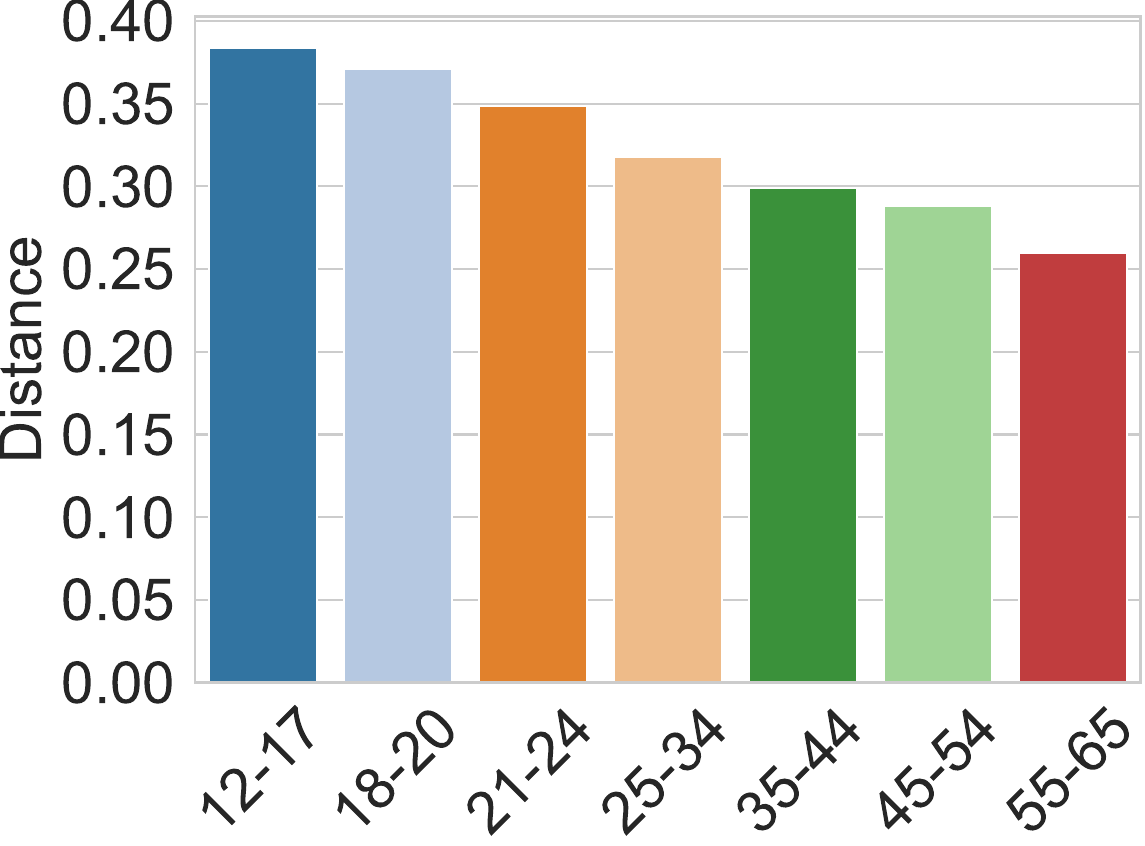}
        \caption{$IGD$ on \acs{bx}}
        \label{fig:results:experiment1:bx:igd}
    \end{subfigure}
    \caption{$AGP$, $GMA$, and $IGD$ across age groups in different datasets.} 
    \label{fig:results:experiment1:agps-igd}
\end{figure*}

\noindent \textbf{Preference Deviation Exploration.} 
\cref{fig:results:experiment1:agps-igd} shows the $AGP$s of age groups and the $APD$ between these. On \ac{ml}, \acl{children} deviate from all other age groups, with an $APD_{\text{\acs{children}}, \text{\acs{mainstream}}} = 0.013$. 
Except for the genres \texttt{Adventure}, \texttt{Horror}, and \texttt{SciFi}, the proportion of all genres is significantly different between \acl{children} and \acl{mainstream}. 
On MLHD, findings are comparable to insights from the music-related dataset, LFM-2b, in the \acl{refWork}. While children's preferences are similar, the older a user gets, the closer their preferences align with those of \acl{mainstream}. Overall, the $APD_{\text{\acs{children}}, \text{\acs{mainstream}}} = 0.0062$, and there are significant differences in the frequencies of all genres except \texttt{Electronic} and \texttt{Rock} between \acl{children} and \acl{mainstream}.

As in MLHD, the older a user gets in \ac{bx}, the closer their preferences align with \acs{mainstream} preferences. However, here, 12-year-olds also stand out as deviating more strongly from children of other ages. For instance, $16$-year-olds' preferences align more closely with \acl{mainstream} than with those of 12-year-olds. Turning to \cref{fig:results:experiment1:bx:genredistribution}, it can be seen that particularly the \texttt{Children} genre becomes markedly more prominent for \acl{children} than any other group. Overall, with $APD_{\text{\acs{children}}, \text{\acs{mainstream}}} = 0.071$, there are significant differences between \acl{children}'s and \acl{mainstream}'s genre consumption of all genres except \texttt{Fantasy/Paranormal}.

\begin{table}[b]
    \centering
    \caption{Results of \textit{popularity extension} per age group$^a$.} 
    \label{tab:results:experiment1:extension}
    \footnotesize

    \begin{tabular}{llrrr}
    \toprule
    & & \textbf{\acl{children}} & \textbf{\acs{mainstream}} & \textbf{\acs{nma}} \\
     \hline \hline \\[-1.8ex]
    \multirow{3}{*}{\rotatebox[]{90}{\textbf{\acs{ml}}}} & \# Interactions/Profile Size & 122.568$^{m}$ & 174.368$^{c, n}$ & 127.021$^{m}$ \\ 
     & Profile Popularity & 0.263$^{m}$ & 0.282$^{c, n}$ & 0.260$^{m}$ \\
     & Profile Age-Popularity & 0.289$^{n}$ & 0.287$^{n}$ & 0.255$^{c, m}$ \\ 
    \hline \\ [-1.8ex]
    \multirow{4}{*}{\rotatebox[]{90}{\textbf{MLHD}}} & \# Interactions & 6341.105$^{m, n}$ & 4758.395$^{c, n}$ & 4174.021$^{c, m}$ \\ 
     & Profile Size & 1244.141$^{m, n}$ & 1461.833$^{c, n}$ & 1814.947$^{c, m}$ \\ 
     & Profile Popularity & 0.079$^{m, n}$ & 0.064$^{c, n}$ & 0.046$^{c, m}$ \\
     & Profile Age-Popularity & 0.067$^{m, n}$ & 0.060$^{c, n}$ & 0.060$^{c, m}$ \\ 
    \hline \\ [-1.8ex]
    \multirow{3}{*}{\rotatebox[]{90}{\textbf{\acs{bx}}}}  & \# Interactions/Profile Size & 4.508$^{m, n}$ & 11.949$^{c}$  & 11.360$^{c}$  \\ 
     & Profile Popularity & 0.057$^{m, n}$ & 0.071$^{c, n}$ & 0.076$^{c, m}$ \\ 
     & Profile Age-Popularity & 0.111$^{m, n}$ & 0.074$^{c, n}$ & 0.081$^{c, m}$ \\ 
    \bottomrule
    \\[-1.8ex]
    \end{tabular}
    \footnotesize{$^a$Significant differences between two groups ($p<0.01$) are annotated with the corresponding pair (\acl{children} ($c$), \acs{mainstream} ($m$), \acp{nma} ($n$)). Note that there are no repeated interactions tracked in \ac{ml} and \ac{bx}. Thus, for these datasets, the number of unique interactions (profile size) corresponds to the number of all interactions.}
\end{table}

\cref{fig:results:experiment1:agps-igd} also shows the $IGD$s across the ages in all three datasets. While no clear differences can be seen between age groups captured by MLHD, results on \ac{ml} show that while children have a comparably high $IGD$, and the youngest \acs{mainstream} users (18-24) have the lowest value, $IGD$ increases the older a user gets. On \ac{bx}, we find \acl{children} have the highest $IGD$, which decreases with age.
Results from the \textit{popularity extension} (\cref{tab:results:experiment1:extension}) show consistent trends between \ac{ml} and \ac{bx}. \texttt{Children} and \acsp{nma} interact with fewer items, on average, than \acl{mainstream}, which are also low in popularity. For \ac{bx}, \acl{children} interact more frequently with items that are popular among their age group.
Contrarily, on MLHD, \acl{children} track more listening events than other age groups, but with fewer distinct songs, i.e., they tend to listen repeatedly to a smaller set of items. Further, the items that they interact with are overall more popular among the entire population.

\noindent \textbf{RS Experiment.} 
Our analysis of the quality of recommendations between age groups when trained on the \texttt{General Set}s of the datasets yields salient differences between datasets and age groups (see \cref{tab:results:Exp2:overview}). On \ac{ml}, for \texttt{MostPop}, \acl{rp3}, and \texttt{iALS}, \acl{mainstream} stand out as a user group that commonly receives better recommendations than \acl{children} or \acsp{nma}: Most performance metrics as well as genre calibration scores are best for this user group. Metric scores for \acl{children} and \acsp{nma} are not significantly different across all metrics and algorithms.
On \ac{bx}, performance metrics differences between \acl{children} and \acl{mainstream} are non-significant across \texttt{MostPop}, \acl{rp3}, and \texttt{iALS}. \acp{nma} stand out with usually significantly worse performance scores. Interestingly, as per $GMC$, \acp{nma} receive the best-aligned recommendations, whereas \acl{children} get the worst across all \acp{ra}. 
For both datasets, \acl{mainstream} stand out as the user group that receives well-aligned recommendations across metrics and \acp{ra}. With $GMC$ on BX being an exception, recommendations for \acl{mainstream} are either significantly better than those of other user groups, or differences to other user groups are non-significant. In other words, \acl{mainstream}, on average, never receive significantly worse recommendations than any other user group. \texttt{Children}'s recommendations are commonly less accurate or less well aligned in terms of $GMC$. 
These insights contradict findings on MLHD. For \texttt{Random}, no significant differences are found (except higher $GMC$ for \acl{children} than for \acl{mainstream}); for the other \acp{ra} \acl{children} receive \textbf{better} recommendations than other age groups, with $GMC$ on \texttt{MostPop} being the exception. In addition, results for metrics between \acl{mainstream} and \acp{nma} are mostly non-significant.

\begin{table*}[t]
    \caption{Average metrics per age group based on the \textit{RS Experiment$^b$.} 
    }
    \label{tab:results:Exp2:overview}
    \setlength{\tabcolsep}{2pt}
\footnotesize
\begin{tabular}{lll|lllll|lllll|lllll}
 &  &  & \multicolumn{5}{c|}{\textbf{ML}} & \multicolumn{5}{c|}{\textbf{MLHD}} & \multicolumn{5}{c}{\textbf{BX}} \\
 & Data & Age Group & 
    nDCG$^\uparrow$ & MRR$^\uparrow$ & MAP$^\uparrow$ & $GMC^\downarrow$ & $PL^{\rightarrow 0}$ &
    nDCG$^\uparrow$ & MRR$^\uparrow$ & MAP$^\uparrow$ & $GMC^\downarrow$ & $PL^{\rightarrow 0}$ &
    nDCG$^\uparrow$ & MRR$^\uparrow$ & MAP$^\uparrow$ & $GMC^\downarrow$ & $PL^{\rightarrow 0}$ \\ \hline
 &  &  &  &  &  &  &  &  &  &  &  &  &  &  \\[-1.8ex]
\multirow{5}{*}{\rotatebox[]{90}{Random}} & Child Set & Children & 
    0.018$^{}$ & 0.039$^{}$ & 0.033$^{}$ & 0.141$^{*}$ & -0.557$^{*}$ &
    0.000 & 0.001 & 0.001 & 0.124$^{*}$ & -0.662$^{*}$ &
    0.004$^{}$ & 0.007$^{}$ & 0.007$^{}$ & 0.261$^{*}$ & 0.165$^{*}$\\
 &  &  &  &  &  &  &  &  &  &  &  &  &  &  \\[-1.8ex]
 & \multirow{3}{*}{General Set} & Children & 
    0.013$^{}$ & 0.024$^{}$ & 0.024$^{}$ & 0.152$^{m,n}$ & -0.685$^{m}$ & 
    0.000 & 0.001 & 0.001 & 0.127$^{m}$ & -0.703$^{m,n}$  &
    0.005$^{m,n}$ & 0.003$^{}$ & 0.003$^{}$ & 0.293$^{m,n}$  & -0.443$^{m,n}$ \\
 &  & Mainstream & 
    0.013$^{n}$ & 0.033$^{n}$ & 0.030$^{n}$ & 0.123$^{c,n}$ & -0.738$^{c,n}$ & 
    0.000 & 0.001 & 0.001 & 0.121$^{c}$ & -0.650$^{c,n}$ &
    0.002$^{c}$ & 0.002$^{}$ & 0.002$^{}$ & 0.212$^{c}$  & -0.547$^{m,n}$ \\
 &  & \Acp{nma} & 
    0.010$^{m}$ & 0.023$^{m}$ & 0.019$^{m}$ & 0.137$^{c,m}$ & -0.698$^{m}$ & 
    0.000 & 0.000 & 0.000 & 0.126 &  -0.596$^{c,m}$ &
    0.001$^{c}$ & 0.002$^{}$ & 0.002$^{}$ & 0.217$^{c}$  & -0.564$^{c}$  \\
 &  &  &  &  &  &  &  &  &  &  &  &  &  &  \\
 
\multirow{5}{*}{\rotatebox[]{90}{MostPop}} & Child Set & Children & 
    0.152$^{*}$ & 0.237$^{}$ & 0.163$^{}$ & 0.140$^{*}$ & 1.410$^{*}$  &
    0.013$^{*}$ & 0.034$^{*}$ & 0.027$^{*}$ & 0.136$^{*}$ & 5.690$^{*}$ &
    0.025$^{}$ & 0.023$^{}$ & 0.020$^{}$ & 0.261$^{*}$  & 3.509$^{*}$  \\
 &  &  &  &  &  &  &  &  &  &  &  &  &  & \\[-1.8ex]
 & \multirow{3}{*}{General Set} & Children & 
    0.129$^{m}$ & 0.208$^{m}$ & 0.146$^{m}$ & 0.151$^{m}$ & 1.768$^{m}$ & 
    0.011$^{m,n}$ & 0.026$^{n}$ & 0.024$^{m,n}$ & 0.139$^{}$ & 6.429$^{m,n}$ &
    0.020$^{}$ & 0.023$^{}$ & 0.021$^{}$ & 0.308$^{m,n}$  & 7.961$^{m,n}$ \\
 &  & Mainstream & 
    0.174$^{c,n}$ & 0.296$^{c,n}$ & 0.196$^{c,n}$ & 0.120$^{c,n}$ & 1.263$^{c,n}$ &
    0.008$^{c}$ & 0.020$^{}$ & 0.018$^{c,n}$ & 0.135$^{n}$ &  7.751$^{c,n}$ &
    0.030$^{n}$ & 0.044$^{n}$ & 0.037$^{n}$ & 0.199$^{c,n}$  &  6.220$^{c}$ \\
 &  & \Acp{nma} & 
    0.126$^{m}$ & 0.213$^{m}$ & 0.147$^{m}$ & 0.147$^{m}$ & 1.594$^{m}$ &
    0.005$^{c}$ & 0.011$^{c}$ & 0.010$^{c,m}$ & 0.146$^{m}$ &  9.229$^{c,m}$ &
    0.021$^{m}$ & 0.025$^{m}$ & 0.022$^{m}$ & 0.181$^{c,m}$  & 5.864$^{c}$ \\
 &  &  &  &  &  &  &  &  &  &  &  &  &  &  \\
 
\multirow{5}{*}{\rotatebox[]{90}{\acl{rp3}}} & Child Set & Children & 
    0.222$^{*}$ & 0.340$^{*}$ & 0.215$^{*}$ & 0.084$^{*}$ & 0.769$^{*}$  &
    0.039$^{*}$ & 0.085$^{}$ & 0.067$^{}$ & 0.059$^{*}$ & 0.709$^{*}$ &
    0.028$^{*}$ & 0.035$^{*}$ & 0.031$^{*}$ & 0.211$^{*}$  & 0.144$^{*}$ \\
 &  &  &  &  &  &  &  &  &  &  &  &  &  &  \\[-1.8ex]
 & \multirow{3}{*}{General Set} & Children & 
    0.287$^{}$ & 0.418$^{m}$ & 0.265$^{m}$ & 0.062$^{}$ &  0.543 &
    0.043$^{m,n}$ & 0.083$^{m,n}$ & 0.069$^{m,n}$ & 0.053$^{m,n}$ &  0.601$^{m,n}$ & 
    0.103$^{n}$ & 0.125$^{}$ & 0.111$^{n}$ & 0.169$^{m,n}$  & 0.379 \\
 &  & Mainstream & 
    0.308$^{n}$ & 0.477$^{c,n}$ & 0.296$^{c,n}$ & 0.059$^{n}$ & 0.593$^{n}$  & 
    0.033$^{c}$ & 0.066$^{c}$ & 0.053$^{c}$ & 0.060$^{c}$ &  0.822$^{c,n}$ &
    0.083$^{n}$ & 0.125$^{n}$ & 0.098$^{n}$ & 0.119$^{c,n}$  & 0.456 \\
 &  & \Acp{nma} & 
    0.281$^{m}$ & 0.435$^{m}$ & 0.277$^{m}$ & 0.063$^{m}$ &  0.542$^{m}$ &
    0.030$^{c}$ & 0.051$^{c}$ & 0.045$^{c}$ & 0.062$^{c}$ & 1.183$^{c,m}$  &
    0.063$^{c,m}$ & 0.083$^{m}$ & 0.067$^{c,m}$ & 0.102$^{c,m}$  & 0.455 \\
 &  &  &  &  &  &  &  &  &  &  &  &  &  & \\
 
\multirow{5}{*}{\rotatebox[]{90}{iALS}} & Child Set & Children & 
    0.197$^{*}$ & 0.334$^{}$ & 0.203$^{*}$ & 0.073$^{*}$ & 0.214$^{*}$  &
    0.033$^{*}$ & 0.067$^{}$ & 0.060$^{}$ & 0.060$^{*}$ & 1.811$^{*}$  &
    0.034$^{*}$ & 0.045$^{*}$ & 0.043$^{*}$ & 0.210$^{*}$ & 0.508$^{*}$ \\  
 &  &  &  &  &  &  &  &  &  &  &  &  &  &  \\[-1.8ex]
 & \multirow{3}{*}{General Set} & Children & 
    0.292$^{m}$ & 0.400$^{m}$ & 0.246$^{m}$ & 0.054$^{m}$ & 0.377$^{m}$  &
    0.038$^{m,n}$ & 0.082$^{m,n}$ & 0.065$^{m,n}$ & 0.042$^{m,n}$  & 0.826$^{m,n}$ & 
    0.106$^{n}$ & 0.123$^{}$ & 0.105$^{}$ & 0.160$^{m,n}$  & 1.242$^{m}$ \\
 &  & Mainstream & 
    0.322$^{c,n}$ & 0.481$^{c,n}$ & 0.295$^{c,n}$ & 0.047$^{c,n}$ & 0.311$^{c,n}$ & 
    0.030$^{c}$ & 0.063$^{c}$ & 0.050$^{c}$ & 0.050$^{c}$ & 1.083$^{c,n}$  &
    0.080$^{n}$ & 0.112$^{n}$ & 0.092$^{n}$ & 0.115$^{c,n}$  & 0.926$^{c}$ \\
 &  & \Acp{nma} & 
    0.302$^{m}$ & 0.449$^{m}$ & 0.272$^{m}$ & 0.055$^{m}$ & 0.363$^{m}$  &
    0.026$^{c}$ & 0.047$^{c}$ & 0.041$^{c}$ & 0.052$^{c}$ &  1.278$^{m,n}$ &
    0.060$^{c,m}$ & 0.081$^{m}$ & 0.069$^{m}$ & 0.096$^{m,n}$  & 1.006 \\\\[-1.8ex]
\end{tabular}
\footnotesize{
$^b$Significant differences between two groups ($p<0.01$) are annotated with the corresponding pair (\acl{children} ($c$), \acs{mainstream} ($m$), \Acp{nma} ($n$)). An asterisk (*) on a \texttt{Child Set} row denotes significant differences in the recommendations for \acl{children} between the \texttt{Child Set} and the \texttt{General Set}. 
}
\end{table*}

Across all datasets \texttt{MostPop}, \acl{rp3}, and \texttt{iALS} lead to a positive popularity lift, i.e., higher average popularity in the recommendations than in the user profiles. Still, there is diverging behavior between MLHD and the other two datasets. On \ac{ml} and \ac{bx}, popularity lift tends to be similar or even higher for \acl{children} than for other user groups. On MLHD, in contrast, the popularity lift for \acl{children} tends to be lower than for \acl{mainstream}. 

Turning our analysis to the differences in recommendation quality for \acl{children} when trained on the \texttt{General Set} versus the \texttt{Child Set}, we observe that for the personalized \acp{ra} training on the \texttt{Child Set} typically leads to \textbf{worse} recommendations on \ac{ml} and \ac{bx}. 
On \texttt{MostPop}, no significant changes are found on these datasets except some improvements in terms of $GMC$ 
when trained on the \texttt{Child Set}, and improved $nDCG$ on \ac{ml}. 

Similarly, on MLHD, differences between $MRR$ and $MAP$ scores are non-significant when trained on either set. However, training on the \texttt{Child Set} leads to \textit{worse} $nDCG$ scores and $GMC$ for \acl{rp3} and \texttt{iALS}. In contrast, recommendation quality increases when trained on the \texttt{Child Set} for \texttt{MostPop}: performance scores are lower; $GMC$ increases.
For all datasets training on the \texttt{Child Set} does not affect performance scores of \texttt{Random}, but it improves the $GMC$.
In terms of popularity, for all \acp{ra} except \texttt{Random}, training on the \texttt{Child Set} consistently leads to lower popularity lift for \acl{children} on \ac{ml} and \ac{bx}, and higher popularity lift on MLHD.

\section{Discussion}
We discuss the obtained results and compare outcomes to the \acl{refWork}, highlighting replicated as well as deviating findings.

\noindent \textbf{Children as Non-Mainstream Users.} 
The \acl{refWork} uncovered differences in preferences, assessed through genre consumption behavior, which this work amplifies, painting a broader picture. There are clear differences in genre consumption across ages in all datasets. Findings on \ac{ml} agree with those in \cite{ungruh2025impact}, which was anticipated given the direct reproduction of the \textit{Preference Deviation Exploration}. Results obtained on MLHD  resemble those on LFM-2b in the \acl{refWork}, i.e., the proportions of genres in $AGP$s align closely;  trends in $APD$ and $IGD$ are similar, highlighting that preference deviations in the original study are reflective of music interactions more generally. Nonetheless, as both datasets are based on interactions from \textit{Last.fm}, key characteristics of users in both datasets will mainly be reflective of the platform's users.
Our analysis on \ac{bx} uncovers findings on a dataset and even a domain unexplored in the \acl{refWork}. Deviations from \acl{children} to the \acs{mainstream} are particularly noticeable here, and younger users show higher preference deviations within their age groups (\cref{fig:results:experiment1:bx:igd}). 

While the key trend of the \acl{refWork}---namely, salient differences in genre preferences between \acl{children} and \acl{mainstream}---is confirmed by our work, the \textit{popularity extension} highlights distinctions between the domains examined. In \ac{ml} and \ac{bx}, children consume fewer items and particularly less popular ones than those consumed by \acl{mainstream}. In the music domain, informed by results from MLHD, children prefer fewer items that are popular in general, which they listen to repeatedly. This finding emphasizes the need for domain-specific considerations when assessing preferences of underrepresented user groups, as generalizations across domains may obscure important nuances in user preferences. 
Differences in preferences highlight that \acl{children} differ in key preferences from adult users. While deviations are even more severe to \acp{nma}, the deviations to \acl{mainstream} highlight a potential oversight of current \ac{rs} research: The datasets used to uncover preferences, interaction patterns, or probe how well a system fares are mainly based on \acl{mainstream} as these are the most prominent user group (see \cref{fig:setup:num_profiles}). However, other user groups such as \acl{children}---the main focus of the \acl{refWork} and this study---may deviate and thus be overlooked.

\noindent \textbf{Dominance of User Groups and Deviating \acs{ra} Behavior.} 
Key differences in behavior between age groups, identified in the \acl{refWork} and confirmed by our findings, raise the question of whether \acp{ra} can capture these distinctions, particularly considering that \acl{mainstream} dominate the data used to train such algorithms. The need to thoroughly explore this concern with the \textit{RS experiment} is exacerbated by the discovery of previously laid out domain-related differences in consumption patterns of \acl{children}.

\ac{ra} performance is directly affected by these differences:
The tendency from the \acl{refWork} that \acl{children} receive recommendations by personalized \acp{ra} that are as good or even better than those of \acl{mainstream} is in line with outcomes of the replicated experiment on MLHD. However, outcomes from the experiments on \ac{ml} and \ac{bx} show that \acl{mainstream} mostly receive accurate and well-aligned recommendations, often to the detriment of recommendation quality for \acl{children} and \acp{nma}.
This contrast may be explained by the tendency of \acl{children} to prefer popular items in the music domain and not in others. On \ac{ml}, \texttt{MostPop} performs significantly worse for \acl{children} than for \acl{mainstream} according to all metrics, and on \ac{bx}, $GMC$ scores are markedly higher for \acl{children}, indicating that genres of popular recommendations do not align with their preferences.  
On MLHD, on the other hand, recommendations by \texttt{MostPop} are, depending on the metric, highly suitable for children, matching outcomes of the \acl{refWork} on LFM-2b.
This preference for popular items on MLHD is reflected by the popularity extension, where \acl{children} exhibit less popularity lift than other user groups on MLHD, but not on others. Due to their preference for already popular items, \acl{children} are impacted less by the popularity bias of the \acp{ra}.

Differences between these datasets highlight the importance of research approaches that acknowledge different user groups in different domains. In domains like music, where \acl{children} prefer popular music and engage with it more intensively, popularity-biased recommenders may incidentally perform better for this group than for others. In contrast, in domains like movies or books, where children prefer less popular items, systems trained on \acs{mainstream} data may struggle to produce equally aligned recommendations for \acl{children}. Good performance in one domain, as highlighted in the \acl{refWork}, does not necessarily translate to another, making multi-domain perspectives crucial if \ac{rs} research truly attempts to acknowledge and serve underrepresented users like \acl{children}.

The \acl{refWork} emphasized inconsistent behavior between \acp{ra}. 
\acl{rp3} leveraged data from other groups to better match \acl{children}'s preferences 
(i.e., \acl{rp3} did not perform as well on \texttt{Child Set}). In contrast, \texttt{iALS} benefited from the focus on \acl{children}'s interactions in the \texttt{Child Set}, yielding similar performance and improved $GMC$~\cite{ungruh2025impact}. Such inconsistencies cannot be found in our study. Instead, \acl{rp3} and \texttt{iALS} both fare significantly less well for \acl{children} when trained on the \texttt{Child Set} across all datasets on most metrics. This may indicate one out of two reasons: First, as assumed by the \acl{refWork}, \acl{children} are indeed `difficult users'~\cite{said2018coherence}, users that are more challenging to recommend to due to deviating preferences; higher $IGD$s (\cref{fig:results:experiment1:agps-igd}), particularly on \ac{ml} and \ac{bx}, support this assumption as such measures can be used to determine difficult users~\cite{said2018coherence, bellogin2011predicting}.
To recommend suitable items to them, an \ac{ra} may require additional information; 
\acp{ra} utilized in our study can leverage \acl{mainstream}' data to create recommendations that align better with what a \acs{children} likes. Second, as the approach developed by the \acl{refWork} to gauge the effect of \acl{mainstream} on recommendations for \acl{children} leads to a reduction of training data available to the \ac{ra}, it reduces the number of interactions that can be leveraged to create fitting recommendations. This may affect \ac{ra} performance negatively.

\texttt{MostPop} leads to no significant differences and, for some metrics, improvement for \acl{children} when training on the \texttt{Child Set}. Recommending items popular \textit{amongst} \acl{children} instead of among all users can be effective to serve this group. 
However, this narrow focus on popular items may have other downsides, potentially reducing fairness, diversity, or novelty in recommendations~\cite{klimashevskaia2024survey, abdollahpouri2021multistakeholder}.

\noindent \textbf{Reproducibility Concerns.} 
We reflect on the changes made to the original setup to enable our experiments.
A majority of datasets used by the \acs{rs} research community does not include age-related information, and if they do, information is coupled with uncertainty: Neither of the datasets used in this study provides entirely accurate information about when age-related data was collected. Thus, the data available serves as the best proxy of users' age.

As in \citet{ungruh2025impact}, we focused on datasets from entertainment domains. In these, we assume that children are mostly `free' to choose what to listen to, read, or watch. We assume that this is reflected by the data available. However, in other domains, where they may have less agency (e.g., e-commerce, education, or tourism), this assumption may not hold. Therefore, the \acl{reference} may be limited to studies in comparable domains.

The datasets selected are the only ones that, to our knowledge, include demographic information; yet it is not always possible to adopt them `as is' to the reference setup.
\ac{bx} and MLHD do not provide genre information, requiring external datasets and APIs for genre annotations. This led to the exclusion of numerous items without reliable genre information. Although we ensured a sufficiently large and balanced set of items across age groups, removing unannotated items raises concerns, particularly as these items may be central to capturing unique or niche preferences.
Further, while genre distributions are suitable to measure preferences in the domains studied, simplifications (such as assuming that artists' genres reflect to each song) may not capture all nuances, and genre equivalents may not exist to sufficiently compare users' preferences in other domains like tourism. 
Properties of some datasets, such as datasets being too large to handle efficiently, non-sequential structure of the data, or unavailable timestamps, limited our ability to fully mimic the original setup.
We adopted alternative strategies grounded in common practices from prior \ac{rs} studies~\cite{kowald2020unfairness, anelli2022top} and recognize that such deviations showcase challenges of consistent evaluation of datasets with differing structures.

\section{Conclusion \& Future Work}
In this work, we reproduced, replicated, and extended the findings of  \citet{ungruh2025impact}, focused on children as a non-mainstream user group. By broadening the analysis across datasets and domains, we confirmed key trends and uncovered new insights into the interplay between children and \ac{rs}. 
Of note, despite \acl{children} being a minority, \acp{ra} can perform well for this user group, and the dominant interactions of the \acs{mainstream} do not necessarily impact children negatively. Yet, this effect is not conclusive across datasets, metrics, and \acp{ra}. In fact, our study spotlights that children are indeed a non-mainstream user group with preferences deviating from those of the \acs{mainstream} for which \acp{ra} may fail to account for. 

Even if a \ac{ra} provides suggestions that align with \acs{children} preferences, this does not mean that the recommendations are necessarily `good' for them. They might still fail in other relevant aspects for \acl{children} such as age-appropriateness~\cite{livingstone2025there}. 
Building on the \acl{refWork}, we provide insights about how current \acp{ra} fare when it comes to \acl{children}. Still, insights are limited by the small number of systems studied, the narrow focus of quality metrics, and a simplified view of children as a homogeneous group. 
Therefore, we see emerging lessons learned as a foundation to move toward \acs{children}-aware \ac{rs}, ones that recognize children as part of the audience and strive to provide recommendations that are not only accurate, but recommend items truly \textit{fitting} to their users. Future research should further leverage the presented extended framework to other non-mainstream user groups (e.g., older adults, niche-interest users, or culturally marginalized communities) to understand how \ac{rs} can better fare for \textit{all} users, not just the statistical majority.

\begin{acks}
This research was supported by the Austrian FFG COMET program. Further, it was supported by grant PID2022-139131NB-I00 funded by MCIN/AEI/10.13039/501100011033.
We thank Michael Ekstrand, Rares Boza, and Gabriel Vigliensoni for general technical support and insights into the datasets. 
\end{acks}

\balance
\bibliographystyle{ACM-Reference-Format}
\bibliography{sample-base}

\end{document}